%% file: main.tex
\documentclass{PoS}
\pdfoutput=1

\newcommand*{\noun}[1]{\textsc{#1}}
\renewcommand*{\emph}[1]{\textit{#1}}

\input{preamble.tex}

\title{
    \hfill{\small{\normalfont\textit{DESY 19-228}}} \\~\\
    Master integrals for all unitarity cuts of massless four-loop propagators}

\ShortTitle{Master integrals for all cuts of massless 4-loop propagators}

\author{\speaker{Vitaly Magerya}\\
        II. Institut f\"ur Theoretische Physik, Universit\"at Hamburg,\\
        Luruper Chaussee 149, 22761 Hamburg, Germany\\
        E-mail: \email{vitaly.magerya@desy.de}}

\author{Andrey Pikelner\\
        Bogoliubov Laboratory of Theoretical Physics, Joint Institute for Nuclear Research,\\
        Joliot-Curie 6, 141980 Dubna, Russia\\
        E-mail: \email{pikelner@theor.jinr.ru}}

\abstract{
Among the unitarity cuts of massless 4-loop propagators two
classes have remained unknown until recently: 2-loop 3-particle
cuts, and 1-loop 4-particle cuts.
In this article we shall discuss the calculation that completes
the master integrals for these cuts: both the motivation and the
methods (including dimensional recurrence relations and direct
integration at higher space-time dimensions).
}

\FullConference{14th International Symposium on Radiative Corrections (RADCOR2019)\\ 
9-13 September 2019\\
Palais des Papes, Avignon, France}

\begin{document}
 
\section{Introduction}

Continuing the effort started last year in~\cite{GMP18}, we have
finished the calculation of all master integrals for unitarity
cuts of massless 4-loop propagators~\cite{MP19}.
Here we would like to briefly explain the motivation behind, and
the method of calculation of these integrals.

\paragraph{Motivation: splitting functions.}
The main motivation for our work are the NNLO corrections to
time-like splitting functions.
As a quick recap: the time-like splitting functions $P_{ab}^{T}(x)$
(for $a,b\in\{q_i,\bar{q}_i,g\}$) act as the kernels in the DGLAP-like
scale evolution equations for the parton fragmentation distributions,
and are related to the space-like splitting functions $P_{ab}^{S}(x)$
that play the same role for the parton distribution functions.
The time-like splitting functions are currently known in the leading color
limit up to N\textsuperscript{3}LO~\cite{MRUV17}; in full they
are known only up to NNLO~\cite{MMV06,MV07,AMV11}, with a caveat:
some of the NNLO terms of $P_{qg}^{T}$ and $P_{gq}^{T}$ are
still undetermined.
The reason for this caveat is that the time-like case was
calculated via an analytic continuation from the space-like case.
At the leading order these two are related simply by
$ P_{ab}^{T} (x)=-x P_{ab}^{S}(1/x),$
but at higher orders this relation does not hold anymore, with
the reason being that, roughly speaking, terms like $\ln(1-1/x)$
that appear in $P_{ab}^{S}(1/x)$ starting at NLO may need to
be continued either as $\ln(1/x-1)+i\pi$ or as $\ln(1/x-1)-i\pi$
depending on the term, which creates an uncertainty, and makes
it impossible to unambiguously restore the full $P_{ab}^{T}$ from just
$P_{ab}^{S}$.
To overcome this uncertainty, a direct calculation of $P_{ab}^{T}(x)$
is required.

At NNLO the ``direct'' calculation of $P_{gq}^{T}$ can be performed~\cite{GM15} by first
calculating the differential cross-section $d\sigma_{p}/dx$ of
\begin{equation}
    \label{eq:eetop}
    e^{+}e^{-}\to\gamma^{*}\to \text{parton }p+\dots,
\end{equation}
in the energy fraction of one of the outgoing partons, 
$ x=2\,q\!\cdot\!k_{p}/q^{2}, $
and then the splitting functions can be read off from the poles of this cross-section,
\begin{equation}
    \frac{d\sigma_{p}}{dx}\sim\dots+\left(\frac{\alpha_{s}}{2\pi}\right)\left\{ \dots-\frac{1}{\epsilon}{P_{pq}^{(0)T}}\right\} +\left(\frac{\alpha_{s}}{2\pi}\right)^{2}\left\{ \dots-\frac{1}{2\epsilon}{P_{pq}^{\left(1\right)T}}\right\} +\left(\frac{\alpha_{s}}{2\pi}\right)^{3}\biggl\{\dots-\frac{1}{3\epsilon}{P_{pq}^{(2)T}}\biggr\},
\end{equation}
where $P_{ab}^{(i)T}(x)$ are the terms of $\alpha_s$ expansion of $P_{ab}^T$.
Note that the needed NNLO correction ${P_{gq}^{(2)T}(x)}$ shows up at $\alpha_{s}^{3}$
of $d\sigma_{g}/dx$, so an {N\textsuperscript{3}LO} calculation is needed.

To obtain $P_{qg}^{(2)T}$ the calculation is the same, except
that in stead of a $\gamma^{*}$, a particle that couples to
gluons is needed: for example a scalar Higgs in an effective
theory with a Higgs-gluon coupling.
Splitting functions are process-independent, so there is some
leeway here.

Of course completing this calculation requires the knowledge the
master integrals for $d\sigma/dx$ at N\textsuperscript{3}LO, and
these are not currently known.
With some preliminary work already done in this direction we
think that the total number of these master integrals is around
500, and that we can calculate them using the differential
equations method (via the $\epsilon$-forms constructed by
Fuchsia~\cite{GM17}), as long as there would be a way to fix
the integration constants that arise in the solution.
Following~\cite{Gituliar15} we propose to do this by exploiting
the observation that an integral over all $x$ values should turn
a differential cross-section into a fully inclusive one,
i.e. $\int dx\,\frac{d\sigma}{dx}(x)=\sigma.$
Applying this observation to each differential master integral,
we can fix its integration constants as long as the corresponding
fully inclusive integral is known.
This brings us to the topic at hand: unitarity cuts of propagators
are exactly the master integrals for the required fully
inclusive cross-sections.
Let us illustrate this point.

\paragraph{Cut integrals.}

For a fully inclusive decay cross-section of an off-shell particle
of momenta~$q$ into~$n$ partons of momenta~$p_i$, the cross-section
is given by
\begin{equation}
    \sigma\sim\sum_{n}\int d\mathrm{PS}_{n}\Big|\langle p_{1},\dots,p_{n}\vert S\vert q\rangle\Big|^{2}=\sum_{n}\int d\mathrm{PS}_{n}\Big|\smallfig{amp1}+\smallfig{amp2}+\dots\Big|^{2},
\end{equation}
where $d\mathrm{PS}_{n}$ is the $n$-particle phase-space element,
\begin{equation}
d\mathrm{PS}_{n}=\left(2\pi\right)^{d}\delta^{d}\!\left(p_{1}+\dots+p_{n}-q\right)\prod_{i=1}^{n}\frac{d^{d}p_{i}}{\left(2\pi\right)^{d}}2\pi\delta\!\left(p_{i}^{2}\right)\Theta\!\left(p_{i}^{0}\right),
\end{equation}
and the graphs stand for Feynman integrals.
Expanding the module squared, each term gives a \emph{cut propagator}, denoted the following way:
\begin{equation}
    \int d\mathrm{PS}_{3}\,\smallfig{amp1}\left(\smallfig{amp2}\right)^{*}=\int d\mathrm{PS}_{3}\,\smallfig{amp1}\;\smallfig{amp2conj}\equiv\smallfig{amp1x2}.
\end{equation}

For N\textsuperscript{3}LO ($\alpha_{s}^{3}$) we need all sets
of cuts of 4-loop propagators: 2-particle cuts, 3-particle, 4-
particle, and 5-particle cuts.
Among these, the 2-particle cuts correspond to 3-loop form-factors,
and have been completed around 2010~\cite{HHM07,HHKS09,LSS10};
the 5-particle cuts are purely phase-space integrals and
where calculated last year~\cite{GMP18}.
Some subset of for 3- and 4-particle cuts have also been calculated
in 2015~\cite{CFHMSS15}, but the majority where unknown until the
full set of the master integrals was completed
in~\cite{MP19}.

\paragraph{Identifying the master integrals.}
To calculate the cut master integrals we first need to identify them.
This is particularly easy if one starts with the
master integrals for the propagators themselves (without cuts):
there are 28 such master integrals, and their values are known
from~\cite{BC10,LSS11}.

To identify the cut master integrals, first, for each of the 28 propagator master integrals one needs
to construct all of its possible cuts.
For example:
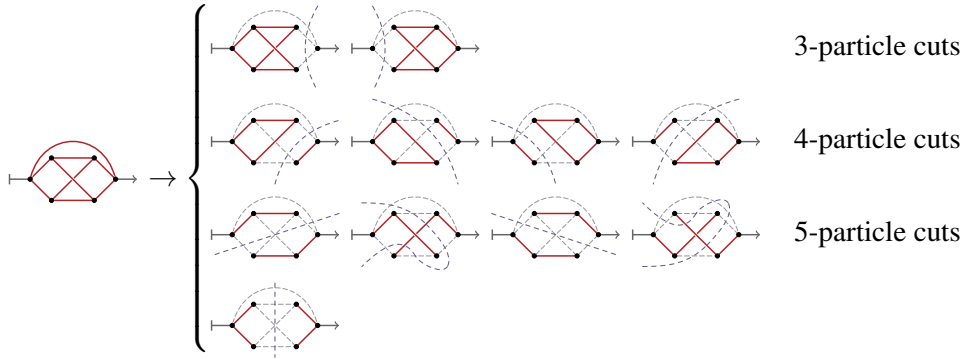
\begin{equation*}
\smallfig{vvvv/12}\to\begin{cases}
\smallfig{vvrr/8a}\;\smallfig{vvrr/8b} & \text{3-particle cuts}\\
\smallfig{vrrr/5a}\;\smallfig{vrrr/5b}\;\smallfig{vrrr/5c}\;\smallfig{vrrr/5d} & \text{4-particle cuts}\\
\smallfig{rrrr/8a}\;\smallfig{rrrr/8b}\;\smallfig{rrrr/8c}\;\smallfig{rrrr/8d} & \text{5-particle cuts}\\
\smallfig{rrrr/2a}
\end{cases}
\end{equation*}

Second, from the resulting set of cuts one needs to remove all the
symmetric duplicate integrals.
For example, all of the following four integrals are identical up to a
complex conjugation:
\begin{equation*}
    \smallfig{vrrr/5a}=\smallfig{vrrr/5b}=\left(\smallfig{vrrr/5c}\right)^{*}=\left(\smallfig{vrrr/5d}\right)^{*}.
\end{equation*}

And this is all, the resulting set will be the final answer. 
There are neither additional IBP relations between the remaining
integrals, nor are there any additional master integrals.

\begin{figure}
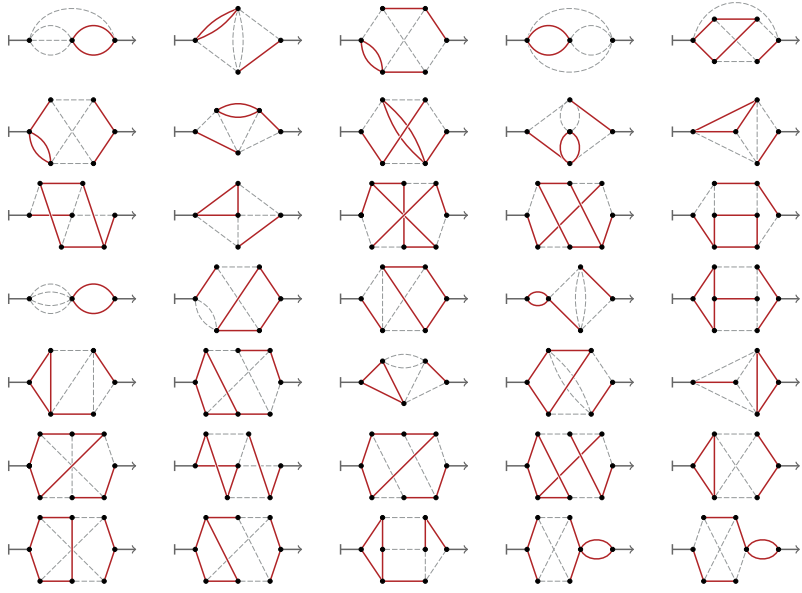

    \label{fig:cut4masters}
    \centering
    \begin{tabular}{ccccc}
    \smallfig{vrrr/1} & \smallfig{vrrr/2} & \smallfig{vrrr/3} & \smallfig{vrrr/4} & \smallfig{vrrr/5}\tabularnewline
    \noalign{\vskip2mm}
    \smallfig{vrrr/6} & \smallfig{vrrr/7} & \smallfig{vrrr/8} & \smallfig{vrrr/9} & \smallfig{vrrr/10}\tabularnewline
    \noalign{\vskip2mm}
    \smallfig{vrrr/11} & \smallfig{vrrr/12} & \smallfig{vrrr/13} & \smallfig{vrrr/14} & \smallfig{vrrr/15}\tabularnewline
    \noalign{\vskip2mm}
    \smallfig{vrrr/16} & \smallfig{vrrr/17} & \smallfig{vrrr/18} & \smallfig{vrrr/19} & \smallfig{vrrr/20}\tabularnewline
    \noalign{\vskip2mm}
    \smallfig{vrrr/21} & \smallfig{vrrr/22} & \smallfig{vrrr/23} & \smallfig{vrrr/24} & \smallfig{vrrr/25}\tabularnewline
    \noalign{\vskip2mm}
    \smallfig{vrrr/26} & \smallfig{vrrr/27} & \smallfig{vrrr/28} & \smallfig{vrrr/29} & \smallfig{vrrr/30}\tabularnewline
    \noalign{\vskip2mm}
    \smallfig{vrrr/31} & \smallfig{vrrr/32} & \smallfig{vrrr/33} & \smallfig{vrrr/34} & \smallfig{vrrr/35}\tabularnewline
    \noalign{\vskip2mm}
    \end{tabular}
    \caption{All 35 master integrals for 4-particle cuts of 4-loop propagators.}
\end{figure}

\begin{figure}
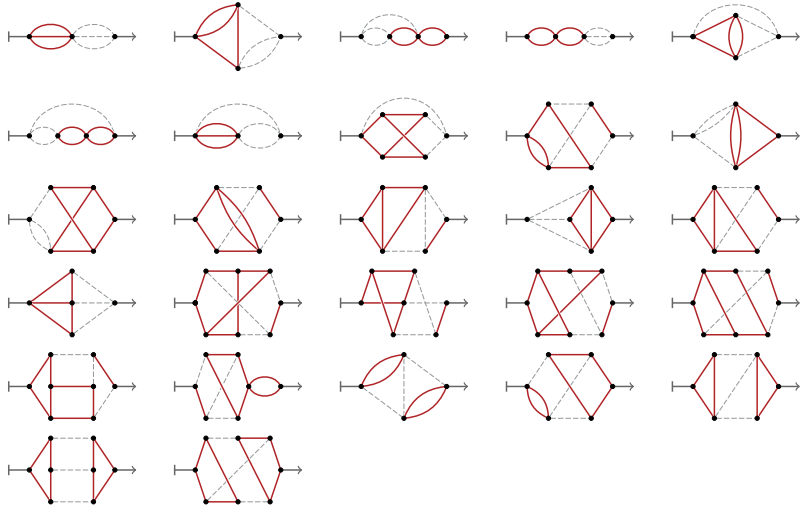

    \label{fig:cut3masters}
    \centering
    \begin{tabular}{ccccc}
    \smallfig{vvrr/1} & \smallfig{vvrr/2} & \smallfig{vvrr/3} & \smallfig{vvrr/4} & \smallfig{vvrr/5}\tabularnewline
    \noalign{\vskip2mm}
    \smallfig{vvrr/6} & \smallfig{vvrr/7} & \smallfig{vvrr/8} & \smallfig{vvrr/9} & \smallfig{vvrr/10}\tabularnewline
    \noalign{\vskip2mm}
    \smallfig{vvrr/11} & \smallfig{vvrr/12} & \smallfig{vvrr/13} & \smallfig{vvrr/14} & \smallfig{vvrr/15}\tabularnewline
    \noalign{\vskip2mm}
    \smallfig{vvrr/16} & \smallfig{vvrr/17} & \smallfig{vvrr/18} & \smallfig{vvrr/19} & \smallfig{vvrr/20}\tabularnewline
    \noalign{\vskip2mm}
    \smallfig{vvrr/21} & \smallfig{vvrr/22} & \smallfig{vrrv/4} & \textrm{\smallfig{vrrv/5}} & \textrm{\smallfig{vrrv/6}}\tabularnewline
    \noalign{\vskip2mm}
    \smallfig{vrrv/7} & \smallfig{vrrv/8} &  &  & \tabularnewline
    \noalign{\vskip2mm}
    \end{tabular}
    \caption{All 27 master integrals for 3-particle cuts of 4-loop propagators.}
\end{figure}

Proceeding this way we have identified the set of master
integrals for 4-particle cuts: 35~integrals in total, depicted in
Table~\ref{fig:cut4masters}; and the same for 3-particle cuts:
27~integrals in total, depicted in Table~\ref{fig:cut3masters}.

\section{Four-particle cut masters via dimensional recurrence relations}

\paragraph{Direct analytic integration?}
Four-particle cut integrals consist of a 4-particle phase space and a 1-loop amplitude.
Calculating these can be as easy as inserting the values of the
1-loop amplitudes (which are known for arbitrary space-time
dimension $d$ in terms of hypergeometric functions), and directly
integrating them over the 4-particle phase-space.
For example knowing the value of a triangle amplitude with all
off-shell legs, one might be able to integrate directly
\begin{equation}
    \smallfig{vrrr/30}=\int\smallfig{triangleoffshell}\frac{1}{s_{13}s_{24}}d\mathrm{PS_{4}}.
\end{equation}

The complication is that 4-particle phase-space has 5 degrees of freedom and a nontrivial shape.
Its parametrisation in terms of the scalar products $s_{ij}=(p_i+p_j)^2/q^2$ looks like
\begin{equation}
    d\mathrm{PS}_4=
        \left(q^{2}\right)^{\frac{3d-4}{2}} 
        \frac{ 2^{4-4d} \pi^{\frac{1}{2}-\frac{3}{2}d} }{ \Gamma\!\left(d-3\right) \Gamma\!\left(\frac{d-1}{2}\right) }
        \left(\Delta_{4}\right)^{\frac{d-5}{2}}
        \Theta\!\left(\Delta_{4}\right) \Theta\!\left(s_{ij}\right) \delta\!\left(1-\sum s_{ij}\right)\prod ds_{ij},
\end{equation}
where $\Delta_{4}$ is the Gram determinant,
\begin{equation}
\Delta_{4}=\det\left|\begin{array}{cccc}
0 & s_{12} & s_{13} & s_{14}\\
s_{12} & 0 & s_{23} & s_{24}\\
s_{13} & s_{23} & 0 & s_{34}\\
s_{14} & s_{24} & s_{34} & 0
\end{array}\right|,
\end{equation}
and the shape of the integration region is given by ${\Theta\!\left(\Delta_{4}\right)\,\Theta(s_{ij})}$.
This is quite a non-trivial shape!
Parametrisations exist that simplify this shape to a hypercube, e.g.
the ``tripole parametrisation''~\cite{GGH03} given by the following change of variables:
\begin{align}
s_{12} & \to\left(1-t\right)(1-y)\left(1-z\right),\\
s_{13} & \to y\left(1-t-v+tv+tvz-2\left(1-2\xi\right)\sqrt{(1-t)(1-v)tvz}\right),\\
s_{14} & \to vy\left(1-z\right),\\
s_{23} & \to z(1-y),\\
s_{24} & \to t(1-y)\left(1-z\right),\\
s_{34} & \to y\left(t-tv+vz-tvz+2\left(1-2\xi\right)\sqrt{(1-t)(1-v)tvz}\right),
\end{align}
but such parametrisations contain square roots, making analytic
integration only possible in the simple cases when the integrand
turns out to be free from these roots.
This is not the case for the majority of the master integrals.
For this reason we turn to a different approach: solving dimensional
recurrence relations.

\paragraph{Dimensional recurrence relations.}
Using a parametric representation of integrals (Feynman for
uncut, Baikov-like for cut) and the IBP tables, we can obtain
dimensional recurrence relations~\cite{Tar96} (DRR) relating
the set master integrals at a raised dimension $I_{i}(d+2)$ to
the initial integrals:
\begin{equation}
    I_{i}\!\left(d+2\right)=M_{ij}\!\left(d\right)I_{j}\!\left(d\right).
\end{equation}

These relations can be solved, and for the 4-particle cuts
specifically, two features of $M_{ij}$ simplify this solution:
\begin{itemize}
\item 
The DRR matrix $M_{ij}$ is triangular (for $j>i$ $M_{ij}=0$).
This means that each $I_i$ can be solved for one by one, with
only the previous $I_{j<i}$ required for the solution---no
coupled blocks need to be considered.
This structure is not surprising, seeing that at most one master
integral is present in each IBP sector for the propagators;
the cuts retain this structure as well.
\item
The diagonal of $M$ is factorizable, with factors linear in $d$:
$M_{ii}=C\prod_{k}\left(d/2-a_{k}\right)^{n_{k}}.$
\end{itemize}

The general solution to DRR has the form of
\begin{equation}
    I_{i}\!\left(d\right)=H_{i}\!\left(d\right)\omega_i\!\left(d\right)+R_{i}\!\left(d\right), 
\end{equation}
where
\begin{itemize}
\item
    $H_{i}$ is a \emph{homogeneous solution}, $H_i(d+2)=M_{ii}(d) H_i(d)$.
    For a factorizable diagonal with no coupled blocks this can immediately be constructed as
\begin{equation}
    \label{eq:hom}
    H_i(d)={C}^{\frac{d}{2}}\prod_{k}\left\{\Gamma^{n_{k}}\!\left(\frac{d}{2}-a_{k}\right)
    \quad\text{or}\quad
    (-1)^{\frac{d}{2}n_k}\Gamma^{-n_{k}}\!\left(a_{k}-\frac{d}{2}+1\right)\right\},
\end{equation}
where both forms of the factors are acceptable, and generally one
chooses one or the other depending on where it would be
most convenient to have the poles and zeros of $H(d)$ reside:
the first form produces poles (or zeros) at $d=2a_k$, $2a_k-2$, $2a_k-4$,
etc., while for the second they are at $2a_k - 2$, $2a_k$, $2a_k+2$, etc.

\item
$R_{i}$ is a \emph{particular solution}. In the absence
of coupled blocks it can be constructed as an infinite sum,
\begin{equation}
    \label{eq:particular}
    R_i(d)=H_i(d)\left\{-\sum^{\infty}_{k=0}\quad\text{or}\quad\sum^{-1}_{k=-\infty}\right\}H^{-1}_i(d+2k+2)\sum_{j<i}M_{ij}(d+2k)\,J_j(d+2k),
\end{equation}
where the direction of summation is chosen based on which one
converges.
The appealing feature of this sum is that it converges geometrically,
and thus can be evaluated numerically with thousands of digits
of precision (as a series in $\epsilon$); its analytic form in
terms of multiple zeta values~\cite{BBV09} (MZVs) can then be
restored using the PSLQ algorithm~\cite{FBA99}.

\item
$\omega$ is an \emph{arbitrary periodic function}, $\omega\!\left(d+2\right)=\omega\!\left(d\right)$.
This function can not be determined from DRR alone;
it plays the same role in the solution of DRR as
the integration constants play in the solution of differential
equations.
\end{itemize}
To find $\omega(d)$ is the main challenge of the method.
Here is an outline of our method (following~\cite{Lee09}):
\begin{enumerate}
\item Find as stripe of width 2 for $d$, such that $I_{i}\!\left(d\right)$ is finite if $\mathrm{Re}\!\left(d\right)\in\left(d_{0},d_{0}+2\right]$, or at least has as few poles as possible.
4-particle cut master integrals $I_{i}$ have surface divergences of UV nature at even~$d$,
but these can be cancelled by factoring out a 1-loop bubble, so in stead of $I_i$ we choose to
work with $J_{i}\equiv I_{i}/\smallfig{vrrr/16}$, which have no UV poles at all, and no IR poles at $d\ge6$.
Thus, we choose the stripe to be $(6,8]$, where $J_i$ are finite.

\item
    Construct the homogeneous solutions $H_i(d)$ with
    eq.~\eqref{eq:hom}, and determine their pole locations. In our
    case these can all be chosen to be finite.
\item
    Construct the particular solutions $R_i(d)$ with
    eq.~\eqref{eq:particular}, and find their pole locations
    by evaluating them numerically for many values of $d$. We use
    \noun{DREAM}~\cite{LM17a} for this evaluation, and all $R_{i}(d)$
    appear to be finite.
\item
    Analyze the behaviour of $\omega_i(d)$ at
    $\mathrm{Im}\,d\to\pm\infty$ by analyzing the same for $J_i$,
    $H_i$, and $R_i$.
    If it can be proven that $|\omega_i(d)| \le K |\mathrm{Im}\,d|^\alpha$
    for some $K$ and $\alpha$, then $\omega_i(d)$ can only have this
    form: $C_{0}+\sum C_{k}\cot\left(\frac{\pi}{2}\left(d-d_{k}\right)\right)$,
    where $C_{k}$ are some constants, and $d_{k}$ are the location of poles of $\omega_i$ (a subset of poles of $J_i$, $H_i$ and $R_i$).
    This is why it is important to choose a stripe for $d$ where
    as few poles appear as possible: the more poles, the more
    constants need to be fixed.

    For smooth $J_{i}$, $H_{i}$, and $R_{i}$ only one choice is
    possible: a constant, $C_{0}$.
\item
    Fix the constants in the ansatz from various considerations.

    Only one constant is needed for us, and we can determine it from the leading pole of $I_i$ at $d$ where it diverges logarithmically in the UV.
    This pole can be calculated by inserting a mass into the loop and looking at the large mass expansion of the result.
    Interestingly, we find that for most integrals (except the trivial ones) the constant
    $C_0$ (and thus $\omega_i$) is zero; a similar situation was already observed previously in~\cite{GMP18} for the
    case of the 5-particle cuts.
\end{enumerate}

Once $\omega_{i}$ are fixed, the masters can be evaluated with
\noun{DREAM} or \noun{SummerTime}~\cite{LM15} numerically to arbitrary precision, and then restored
in terms of MZVs via PSLQ.
Here is an example result:
\begin{align}
\smallfig{vrrr/24}= & \frac{\left(\smallfig{vrrr/16}\right)^{*}}{\left(q^{2}\right)^{3}}\Big[-6\zeta_{2}\,\frac{1}{\epsilon^{2}}+\Big(59\zeta_{2}-60\zeta_{3}\Big)\,\frac{1}{\epsilon}+\Big(-203\zeta_{2}+590\zeta_{3}-\\
 & -129\zeta_{2}^{2}\Big)+\Big(288\zeta_{2}-2030\zeta_{3}+\frac{2537}{2}\zeta_{2}^{2}+192\zeta_{2}\zeta_{3}-1806\zeta_{5}\Big)\,\epsilon+\mathcal{O}\!\left(\epsilon^{2}\right)\Big]. \nonumber
\end{align}

Overall, we've restored the series up to MZVs of weight 12. They have
poles up to $1/\epsilon^{5}$, and zetas up to weight~6 in the $\epsilon$-finite
part.

To cross-check these results we have evaluated them numerically
via simple Monte-Carlo (without e.g. sector decomposition) by
combining Feynman parametrisation of the 1-loop amplitude with
the tripole parametrisation of $d\mathrm{PS}_4$.
This is possible without running into divergences because
UV divergence only comes from the pre-factors of the Feynman
parametrisation (which can be factorized), and evaluating the
integrals in $d=6-2\epsilon$ removes the IR divergences, so the
result is finite.

\section{Three-particle cut masters via integration in 6 dimensions}
\label{sec:3cut}

Solving DRR for 3-particle cuts is much more difficult than for
the 4-particle ones because they contain 2-loop amplitudes,
and thus many more UV poles from subdivergences, so the
ans\"atze for $\omega_i$ may have dozens of free parameters, and we
don't have enough information to fix them (yet).

On the flip side, the 3-particle phase space is simpler that the
4-particle one, and we are fully able to represent these
master integrals as a 2-loop 1$\to$3 amplitude and 3-particle
phase space, e.g.
\begin{equation}
    \smallfig{vvrr/8}=\int\smallfig{1to3/na19}d\mathrm{PS_{3}},
\end{equation}
parameterize the phase-space via $s_{ij}=\left(p_{i}+p_{j}\right)^{2}/q^{2}$ as
\begin{equation}
    \label{eq:dps3}
    d\mathrm{PS}_{3}=\left(q^{2}\right)^{d-3}\frac{2^{4-3d}\pi^{\frac{3}{2}-d}}{\Gamma\!\left(\frac{d-2}{2}\right)\Gamma\!\left(\frac{d-1} {2}\right)}\left(s_{12}s_{13}s_{23}\right)^{\frac{d-4}{2}}\delta\! \left(1-\sum s_{ij}\right)\prod ds_{ij},
\end{equation}
and integrate the result---as long as we know the values of the
2-loop 1$\to$3 amplitudes---and we do: the master integrals are known
from~\cite{GR00,GR01} as series in $\epsilon$ with coefficients
being multiple polylogarithms~\cite{Goncharov98} (depending on $s_{12}$ and $s_{13}$)
up to transcendental weight~4.

There are two caveats here.
The first one is that just taking a 2-loop 1$\to$3 amplitudes as
series in $\epsilon$, multiplying it by the expansion of $d\mathrm{PS}_3$,
and integrating order by order in $\epsilon$
will give infinity in each order:
\begin{equation}
    \int\smallfig{1to3/na19}d\mathrm{PS_{3}}=\int\frac{C}{s_{12}^{2}}\left(\frac{1}{\epsilon^{4}}-\frac{2\ln s_{12}}{\epsilon^{3}}+\dots\right)d\mathrm{PS_{3}}=\frac{\infty}{\epsilon^{4}}+\frac{\infty}{\epsilon^{3}}+\dots.
\end{equation}

The reason is that the integrals are IR-divergent, and for them taking a series in $\epsilon$ does not commute with integration, so the integral must be taken before the expansion into series.

The solution is to switch to an IR-finite basis of integrals.
Any such basis should work, but a simple choice is to take
the same list of master integrals only evaluated at
$d=6-2\epsilon$.
This is a dimension high enough to cancel the IR poles, so the
integration can be done order by order in~$\epsilon$.

The second caveat is that because the $\epsilon$-finite part of
the propagator master integrals contain MZVs up to weight~7,
and the same for 4-particle cuts go up to weight~6 (as
we've already seen), it means that we need to know the 2-loop
1$\to$3 amplitudes up to at least weight~6 too---but the results
in~\cite{GR00,GR01} only go to weight~4.
For this reason we have recomputed the 1$\to$3 master integrals
up to weight~7, by:
\begin{itemize}
    \item writing down differential equation systems for them in $s_{12}$ and $s_{23}$;
    \item finding a change of basis that simultaneously reduces both systems into an $\epsilon$-form~\cite{Henn13,Lee14} using Fuchsia\footnote{The version of Fuchsia available at \href{https://github.com/magv/fuchsia.cpp}{github.com/magv/fuchsia.cpp} does this automatically.}~\cite{GM17};
    \item constructing the general solution, containing a
    mixture of multiple polylogarithms in the form of
    $\sum R\!\left(s_{ij},C\right)G\!\left(\{0,1,1-s_{13},-s_{13}\};s_{12}\right)G\!\left(\{0,1\};s_{13}\right)$,
    where $R$ are some rational functions,  $C$ are the integration
    constants, and $G$ are the multiple polylogarithms;
    \item fixing the integration constants $C$ by matching with
    the known single-scale integrals, enforcing regularity at
    $s_{ij}\to1$ (because the integrals are massless), and
    enforcing regularity at $s_{ij}\to0$ for planar integrals
    if $i$ and $j$ are not adjacent.
\end{itemize}

With the 2-loop 1$\to$3 amplitudes obtained this way up to
weight~7, we can proceed to evaluate the 3-particle cut integrals
to the same weight.
For this:
\begin{enumerate}
\item Use dimensional recurrence for the 1$\to$3 amplitudes to get them
as series around $d=6-2\epsilon$.
\item Multiply that by the $\epsilon$-expansion of $d\mathrm{PS}_{3}$ (in
$d=6-2\epsilon$), and integrate order by order.
\item Use dimensional recurrence for the resulting integrals to lower the series
back to $d=4-2\epsilon$.
\end{enumerate}

Overall, this way we get the results as series in $\epsilon$ up
to MZVs of weight 7, with poles up to $1/\epsilon^{6}$, and zetas
up to weight~6 in the $\epsilon$-finite part. As an example:
\begin{align*}
\smallfig{vvrr/12}= & \frac{\smallfig{vvrr/4}}{\left(q^{2}\right)^{2}}\Big[2\zeta_{2}\,\frac{1}{\epsilon}+\Big(-13\zeta_{2}+16\zeta_{3}\Big)+\Big(27\zeta_{2}-104\zeta_{3}+\frac{156}{5}\zeta_{2}^{2}\Big)\,\epsilon+\\
 & +\Big(-18\zeta_{2}+216\zeta_{3}-\frac{1014}{5}\zeta_{2}^{2}-90\zeta_{2}\zeta_{3}+448\zeta_{5}\Big)\,\epsilon^{2}+\mathcal{O}\!\left(\epsilon^{3}\right)\Big].
\end{align*}

To cross-check the results we have compared them to the few
integrals known from~\cite{CFHMSS15}, and also by evaluating them
numerically using \noun{FIESTA}~\cite{Smirnov15}. To achieve the
latter one needs to find a parametrisation for the cut integrals
suitable for sector decomposition; in the case of 3-particle
cuts this can be done by combining Feynman parameterization for
the loop part, and eq.~\eqref{eq:dps3} for the phase space.

\section{Cross-check via Cutkosky relations}

Cutkosky relations connect virtual integrals to their cuts; now
that we have all cuts of 4-loop propagators available, these can
act as the final consistency check. To write them down, for any
Feynman diagram $F$:
\begin{equation}
    F+F^{*}=-\sum_{i}\mathrm{Cut}_{i}F.
\end{equation}

This simple form is valid for Feynman diagrams, so to use it
our integrals must be dressed in Feynman rules; any consistent
set would do, and the simplest one comes from a scalar $\phi^n$
theory:
\begin{equation}
    \smallfig{3pt}=\smallfig{4pt}=\smallfig{5pt}=\dots=i,
    \hspace{2em}
    \smallfig{propagator}=\frac{i}{p^{2}+i0},
    \hspace{2em}
    \smallfig{cut-propagator}=2\pi\delta^{+}\!\left(p^{2}\right).
\end{equation}

Combining the above, we get one relation for each of the 31 propagator masters:
\begin{align}
2\,\mathrm{Im}\,\smallfig{vvvv/12} & =2\,\mathrm{Re}\,\smallfig{vvrr/8}+4\,\mathrm{Im}\,\smallfig{vrrr/5}-4\,\smallfig{rrrr/8}-\smallfig{rrrr/2}\\
2\,\mathrm{Im}\,\smallfig{vvvv/14} & =2\,\mathrm{Re}\,\smallfig{vvrr/16}+4\,\mathrm{Im}\,\smallfig{vrrr/12}-2\,\smallfig{rrrr/17}\\
 & \ldots\nonumber
\end{align}

We now have most of the integrals up to weight~12 (with only
the 3-particle cuts up to weight~7) inserting these values into
these equations gives us our cross-check.

\section{Three-particle cuts via dimensional recurrence relations}

We can additionally use Cutkosky relations to obtain
information about one of the cuts from the knowledge of the
others, in particular to obtain 3-particle cuts to weight~12.
Two cases need to be considered:
\begin{enumerate}
\item The {easy case}: only one 3-particle cut enters a Cutkosky relation. For example:
\begin{equation}
    \underbrace{2\,\mathrm{Im}\,\smallfig{vvvv/13}=I\,\smallfig{vvvr/10}-\smallfig{rrrr/10}}_{\mathclap{\text{known to }\zeta_{12}}}-\underbrace{I\,\smallfig{vrrr/2}}_{\mathclap{\text{just calculated}}}+\underbrace{\smallfig{vvrr/10}}_{\mathclap{\text{3-particle cut}}}.
\end{equation}
\item The {hard case}: multiple 3-particle cuts enter a Cutkosky
relation. For example:
\begin{align*}
2\,\mathrm{Im}\,\smallfig{vvvv/23}= & -2I\,\smallfig{vrrr/3}-I\,\smallfig{vrrr/6}+I\,\smallfig{vrrr/17}-2\,\smallfig{rrrr/27}\\
 & +I\,\smallfig{vvvr/12}+\underbrace{2\,\smallfig{vvrr/9}+\smallfig{vvrr/11}-\smallfig{vrrv/5}}_{\mathclap{\text{3-particle cuts}}}.
\end{align*}

In this case only the sum is constrained. It turns out that this
constrain on the sum, together with a few leading terms of the
$\epsilon$-expansion that we have calculated in Section~\ref{sec:3cut}
gives us enough information to fix all the constants in the
ansatz for $\omega_i(d)$ of the DRR solution for 3-particle cut
master integrals, and thus solve the DRR for them completely.

This upgrades our knowledge of the 3-particle cuts from weight~7
to weight~12 (and more).
\end{enumerate}

\section{Summary}
We have calculated the full set of master integrals for 3- and
4-particle unitarity cuts of massless 4-loop propagators, thus
completing the knowledge of all such cuts.
Both direct phase-space integration and the solution of dimensional
recurrence relations where used for this.
As a byproduct we have also re-calculated the 2-loop 1$\to$3
master integrals to transcendental weight~7 (an upgrade from
weight-4 result from \cite{GR00,GR01}).
Further research includes using these integrals as boundary conditions for the differential master integrals of the same form, and using those for the calculations of NNLO splitting functions.

\bibliographystyle{JHEPmod}
\bibliography{main}

\end{document}

%% file: preamble.tex
\usepackage{tikz}
\usepackage{graphicx}
\usepackage{bbding}
\usepackage{mathtools}
\usepackage{microtype}
\usepackage{slashed}
\usepackage{shuffle}
\usepackage{longtable}
\usepackage{bm}

\definecolor{SapGreen}{HTML}{3B7B3B}
\definecolor{PhtaloGreen}{HTML}{2b7a76}
\definecolor{EmeraldGreen}{HTML}{1ea78d}
\definecolor{EnglishRed}{HTML}{b02427}
\definecolor{IronOxideRed}{HTML}{a13634}
\definecolor{CadmiumRed}{HTML}{df2b3f}
\definecolor{Vermillion}{HTML}{e14235}

\usepackage{dsfont}

\ifcsname pdfmdfivesum\endcsname
\else
    \ifcsname mdfivesum\endcsname
    \else
        \errmessage{Neither pdfmdfivesum nor mdfivesum commands are present. Fail}
    \fi
\fi

\usetikzlibrary{calc}
\usetikzlibrary{external}
\usetikzlibrary{fit}
\usetikzlibrary{arrows.meta}
\tikzset{baseline=-0.6ex}
\tikzset{inner sep=0}

\pgfdeclarelayer{nodelayer}
\pgfdeclarelayer{edgelayer}
\pgfsetlayers{edgelayer,nodelayer,main}
\tikzstyle{none}=[]
\input{all.tikzstyles}

\newcommand*{\smallfig}[1]{\raisebox{0.15ex}{\scalebox{0.75}{%
    \scalebox{0.75}{\input{fig/#1.tikz}}%
}}}

%% file: all.tikzstyles
\tikzstyle{dot}=[fill=black, shape=circle, draw=black, inner sep=1pt]
\tikzstyle{blob}=[fill=white, draw={rgb,255: red,176; green,36; blue,39}, dashed, shape=circle, line width=1pt, inner sep=5pt]
\tikzstyle{align east}=[anchor=east]
\tikzstyle{align west}=[anchor=west]

\tikzstyle{edge}=[-, draw={rgb,255: red,176; green,36; blue,39}, line width=1pt, preaction={{draw=white,line width=2pt}}, line cap=rect]
\tikzstyle{massive edge}=[-, draw={rgb,255: red,133; green,119; blue,181}, line width=2.5pt, preaction={{draw=white,line width=3pt}}, line cap=rect]
\tikzstyle{cut edge}=[-, draw={rgb,255: red,147; green,151; blue,152}, line width=0.5pt, densely dashed, line cap=rect]
\tikzstyle{incoming edge}=[line width=1pt, line cap=rect, draw={rgb,255: red,102; green,102; blue,102}, {|-}]
\tikzstyle{outgoing edge}=[line width=1pt, line cap=rect, draw={rgb,255: red,102; green,102; blue,102}, -{Classical TikZ Rightarrow[]}]
\tikzstyle{cut}=[-, draw={rgb,255: red,104; green,89; blue,145}, line width=0.7pt, dashed, line cap=round]

%% file: main.bbl
\providecommand{\href}[2]{#2}\begingroup\raggedright\begin{thebibliography}{10}

\bibitem{GMP18}
O.~Gituliar, V.~Magerya and A.~Pikelner, \emph{{Five-Particle Phase-Space
  Integrals in QCD}},
  \href{https://doi.org/10.1007/JHEP06(2018)099}{\emph{JHEP} {\bfseries 06}
  (2018) 099} [\href{https://arxiv.org/abs/1803.09084}{{\ttfamily
  1803.09084}}].

\bibitem{MP19}
V.~Magerya and A.~Pikelner, \emph{{Cutting massless four-loop propagators}},
  \href{https://doi.org/10.1007/JHEP12(2019)026}{\emph{JHEP} {\bfseries 12}
  (2019) 026} [\href{https://arxiv.org/abs/1910.07522}{{\ttfamily
  1910.07522}}].

\bibitem{MRUV17}
S.~Moch, B.~Ruijl, T.~Ueda, J.~A.~M. Vermaseren and A.~Vogt, \emph{{Four-Loop
  Non-Singlet Splitting Functions in the Planar Limit and Beyond}},
  \href{https://doi.org/10.1007/JHEP10(2017)041}{\emph{JHEP} {\bfseries 10}
  (2017) 041} [\href{https://arxiv.org/abs/1707.08315}{{\ttfamily
  1707.08315}}].

\bibitem{MMV06}
A.~Mitov, S.~Moch and A.~Vogt, \emph{{Next-to-Next-to-Leading Order Evolution
  of Non-Singlet Fragmentation Functions}},
  \href{https://doi.org/10.1016/j.physletb.2006.05.005}{\emph{Phys. Lett.}
  {\bfseries B638} (2006) 61}
  [\href{https://arxiv.org/abs/hep-ph/0604053}{{\ttfamily hep-ph/0604053}}].

\bibitem{MV07}
S.~Moch and A.~Vogt, \emph{{On third-order timelike splitting functions and
  top-mediated Higgs decay into hadrons}},
  \href{https://doi.org/10.1016/j.physletb.2007.10.069}{\emph{Phys. Lett.}
  {\bfseries B659} (2008) 290}
  [\href{https://arxiv.org/abs/0709.3899}{{\ttfamily 0709.3899}}].

\bibitem{AMV11}
A.~A. Almasy, S.~Moch and A.~Vogt, \emph{{On the Next-to-Next-to-Leading Order
  Evolution of Flavour-Singlet Fragmentation Functions}},
  \href{https://doi.org/10.1016/j.nuclphysb.2011.08.028}{\emph{Nucl. Phys.}
  {\bfseries B854} (2012) 133}
  [\href{https://arxiv.org/abs/1107.2263}{{\ttfamily 1107.2263}}].

\bibitem{GM15}
O.~Gituliar and S.~Moch, \emph{{Towards three-loop QCD corrections to the
  time-like splitting functions}},
  \href{https://doi.org/10.5506/APhysPolB.46.1279}{\emph{Acta Phys. Polon.}
  {\bfseries B46} (2015) 1279}
  [\href{https://arxiv.org/abs/1505.02901}{{\ttfamily 1505.02901}}].

\bibitem{GM17}
O.~Gituliar and V.~Magerya, \emph{{Fuchsia: a tool for reducing differential
  equations for Feynman master integrals to epsilon form}},
  \href{https://doi.org/10.1016/j.cpc.2017.05.004}{\emph{Comput. Phys. Commun.}
  {\bfseries 219} (2017) 329}
  [\href{https://arxiv.org/abs/1701.04269}{{\ttfamily 1701.04269}}].

\bibitem{Gituliar15}
O.~Gituliar, \emph{{Master integrals for splitting functions from differential
  equations in QCD}},
  \href{https://doi.org/10.1007/JHEP02(2016)017}{\emph{JHEP} {\bfseries 02}
  (2016) 017} [\href{https://arxiv.org/abs/1512.02045}{{\ttfamily
  1512.02045}}].

\bibitem{HHM07}
G.~Heinrich, T.~Huber and D.~{Ma\^itre}, \emph{{Master integrals for fermionic
  contributions to massless three-loop form-factors}},
  \href{https://doi.org/10.1016/j.physletb.2008.03.028}{\emph{Phys. Lett.}
  {\bfseries B662} (2008) 344}
  [\href{https://arxiv.org/abs/0711.3590}{{\ttfamily 0711.3590}}].

\bibitem{HHKS09}
G.~Heinrich, T.~Huber, D.~A. Kosower and V.~A. Smirnov, \emph{{Nine-Propagator
  Master Integrals for Massless Three-Loop Form Factors}},
  \href{https://doi.org/10.1016/j.physletb.2009.06.038}{\emph{Phys. Lett.}
  {\bfseries B678} (2009) 359}
  [\href{https://arxiv.org/abs/0902.3512}{{\ttfamily 0902.3512}}].

\bibitem{LSS10}
R.~N. Lee, A.~V. Smirnov and V.~A. Smirnov, \emph{{Analytic Results for
  Massless Three-Loop Form Factors}},
  \href{https://doi.org/10.1007/JHEP04(2010)020}{\emph{JHEP} {\bfseries 04}
  (2010) 020} [\href{https://arxiv.org/abs/1001.2887}{{\ttfamily 1001.2887}}].

\bibitem{CFHMSS15}
M.~Czakon, P.~Fiedler, T.~Huber, M.~Misiak, T.~Schutzmeier and M.~Steinhauser,
  \emph{{The~$(Q_{7}, Q_{1,2})$ contribution to $\overline{B}\to{X}_s\gamma$ at
  $\mathcal{O}\left({\alpha}_{\mathrm{s}}^2\right)$}},
  \href{https://doi.org/10.1007/JHEP04(2015)168}{\emph{JHEP} {\bfseries 04}
  (2015) 168} [\href{https://arxiv.org/abs/1503.01791}{{\ttfamily
  1503.01791}}].

\bibitem{BC10}
P.~A. Baikov and K.~G. Chetyrkin, \emph{{Four Loop Massless Propagators: An
  Algebraic Evaluation of All Master Integrals}},
  \href{https://doi.org/10.1016/j.nuclphysb.2010.05.004}{\emph{Nucl. Phys.}
  {\bfseries B837} (2010) 186}
  [\href{https://arxiv.org/abs/1004.1153}{{\ttfamily 1004.1153}}].

\bibitem{LSS11}
R.~N. Lee, A.~V. Smirnov and V.~A. Smirnov, \emph{{Master Integrals for
  Four-Loop Massless Propagators up to Transcendentality Weight Twelve}},
  \href{https://doi.org/10.1016/j.nuclphysb.2011.11.005}{\emph{Nucl. Phys.}
  {\bfseries B856} (2012) 95}
  [\href{https://arxiv.org/abs/1108.0732}{{\ttfamily 1108.0732}}].

\bibitem{GGH03}
A.~{Gehrmann-De~Ridder}, T.~Gehrmann and G.~Heinrich, \emph{{Four particle
  phase space integrals in massless QCD}},
  \href{https://doi.org/10.1016/j.nuclphysb.2004.01.023}{\emph{Nucl.Phys.}
  {\bfseries B682} (2004) 265}
  [\href{https://arxiv.org/abs/hep-ph/0311276}{{\ttfamily hep-ph/0311276}}].

\bibitem{Tar96}
O.~V. Tarasov, \emph{{Connection between Feynman integrals having different
  values of the space-time dimension}},
  \href{https://doi.org/10.1103/PhysRevD.54.6479}{\emph{Phys. Rev.} {\bfseries
  D54} (1996) 6479} [\href{https://arxiv.org/abs/hep-th/9606018}{{\ttfamily
  hep-th/9606018}}].

\bibitem{BBV09}
J.~{Bl\"umlein}, D.~J. Broadhurst and J.~A.~M. Vermaseren, \emph{{The Multiple
  Zeta Value Data Mine}},
  \href{https://doi.org/10.1016/j.cpc.2009.11.007}{\emph{Comput. Phys. Commun.}
  {\bfseries 181} (2010) 582}
  [\href{https://arxiv.org/abs/0907.2557}{{\ttfamily 0907.2557}}].

\bibitem{FBA99}
H.~Ferguson, D.~Bailey and S.~Arno, \emph{Analysis of {PSLQ}, an integer
  relation finding algorithm},
  \href{https://doi.org/10.1090/S0025-5718-99-00995-3}{\emph{Mathematics of
  Computation of the American Mathematical Society} {\bfseries 68} (1999) 351}.

\bibitem{Lee09}
R.~N. Lee, \emph{{Space-time dimensionality D as complex variable: Calculating
  loop integrals using dimensional recurrence relation and analytical
  properties with respect to D}},
  \href{https://doi.org/10.1016/j.nuclphysb.2009.12.025}{\emph{Nucl. Phys.}
  {\bfseries B830} (2010) 474}
  [\href{https://arxiv.org/abs/0911.0252}{{\ttfamily 0911.0252}}].

\bibitem{LM17a}
R.~N. Lee and K.~T. Mingulov, \emph{{DREAM, a program for arbitrary-precision
  computation of dimensional recurrence relations solutions, and its
  applications}},  \href{https://arxiv.org/abs/1712.05173}{{\ttfamily
  1712.05173}}.

\bibitem{LM15}
R.~N. Lee and K.~T. Mingulov, \emph{{Introducing SummerTime: a package for
  high-precision computation of sums appearing in DRA method}},
  \href{https://doi.org/10.1016/j.cpc.2016.02.018}{\emph{Comput. Phys. Commun.}
  {\bfseries 203} (2016) 255}
  [\href{https://arxiv.org/abs/1507.04256}{{\ttfamily 1507.04256}}].

\bibitem{GR00}
T.~Gehrmann and E.~Remiddi, \emph{{Two loop master integrals for $\gamma^*$
  $\to$ 3 jets: The~Planar topologies}},
  \href{https://doi.org/10.1016/S0550-3213(01)00057-8}{\emph{Nucl. Phys.}
  {\bfseries B601} (2001) 248}
  [\href{https://arxiv.org/abs/hep-ph/0008287}{{\ttfamily hep-ph/0008287}}].

\bibitem{GR01}
T.~Gehrmann and E.~Remiddi, \emph{{Two loop master integrals for $\gamma^*$
  $\to$ 3 jets: The~Nonplanar topologies}},
  \href{https://doi.org/10.1016/S0550-3213(01)00074-8}{\emph{Nucl. Phys.}
  {\bfseries B601} (2001) 287}
  [\href{https://arxiv.org/abs/hep-ph/0101124}{{\ttfamily hep-ph/0101124}}].

\bibitem{Goncharov98}
A.~B. Goncharov, \emph{{Multiple polylogarithms, cyclotomy and modular
  complexes}}, \href{https://doi.org/10.4310/MRL.1998.v5.n4.a7}{\emph{Math.
  Res. Lett.} {\bfseries 5} (1998) 497}
  [\href{https://arxiv.org/abs/1105.2076}{{\ttfamily 1105.2076}}].

\bibitem{Henn13}
J.~M. Henn, \emph{{Multiloop integrals in dimensional regularization made
  simple}}, \href{https://doi.org/10.1103/PhysRevLett.110.251601}{\emph{Phys.
  Rev. Lett.} {\bfseries 110} (2013) 251601}
  [\href{https://arxiv.org/abs/1304.1806}{{\ttfamily 1304.1806}}].

\bibitem{Lee14}
R.~N. Lee, \emph{{Reducing differential equations for multiloop master
  integrals}}, \href{https://doi.org/10.1007/JHEP04(2015)108}{\emph{JHEP}
  {\bfseries 04} (2015) 108} [\href{https://arxiv.org/abs/1411.0911}{{\ttfamily
  1411.0911}}].

\bibitem{Smirnov15}
A.~V. Smirnov, \emph{{FIESTA4: Optimized Feynman integral calculations with GPU
  support}}, \href{https://doi.org/10.1016/j.cpc.2016.03.013}{\emph{Comput.
  Phys. Commun.} {\bfseries 204} (2016) 189}
  [\href{https://arxiv.org/abs/1511.03614}{{\ttfamily 1511.03614}}].

\end{thebibliography}\endgroup
